\documentclass[dvipdfmx]{aastex63}

\usepackage{color}
\usepackage[normalem]{ulem}
\usepackage{multirow}
\usepackage{graphicx}
\received{}
\revised{}
\accepted{}
\submitjournal{ApJ}

\shorttitle{An ALMA-ACA CO Survey in the SMC North}
\shortauthors{Tokuda et al.}

\begin{document}

\title{An Unbiased CO Survey Toward the Northern Region of the Small Magellanic Cloud with the Atacama Compact Array. I. Overview: CO Cloud Distributions}

\correspondingauthor{Kazuki Tokuda}
\email{tokuda@p.s.osakafu-u.ac.jp}

\author[0000-0002-2062-1600]{Kazuki Tokuda}
\affiliation{Department of Physical Science, Graduate School of Science, Osaka Prefecture University, 1-1 Gakuen-cho, Naka-ku, Sakai, Osaka 599-8531, Japan}
\affiliation{National Astronomical Observatory of Japan, National Institutes of Natural Science, 2-21-1 Osawa, Mitaka, Tokyo 181-8588, Japan}

\author{Hiroshi Kondo}
\affiliation{Department of Physical Science, Graduate School of Science, Osaka Prefecture University, 1-1 Gakuen-cho, Naka-ku, Sakai, Osaka 599-8531, Japan}

\author{Takahiro Ohno}
\affiliation{Department of Physics, Nagoya University, Chikusa-ku, Nagoya 464-8602, Japan}

\author{Ayu Konishi}
\affiliation{Department of Physical Science, Graduate School of Science, Osaka Prefecture University, 1-1 Gakuen-cho, Naka-ku, Sakai, Osaka 599-8531, Japan}

\author{Hidetoshi Sano}
\affiliation{National Astronomical Observatory of Japan, National Institutes of Natural Science, 2-21-1 Osawa, Mitaka, Tokyo 181-8588, Japan}

\author{Kisetsu Tsuge}
\affiliation{Department of Physics, Nagoya University, Chikusa-ku, Nagoya 464-8602, Japan}

\author[0000-0001-6149-1278]{Sarolta Zahorecz}
\affiliation{Department of Physical Science, Graduate School of Science, Osaka Prefecture University, 1-1 Gakuen-cho, Naka-ku, Sakai, Osaka 599-8531, Japan}
\affiliation{National Astronomical Observatory of Japan, National Institutes of Natural Science, 2-21-1 Osawa, Mitaka, Tokyo 181-8588, Japan}

\author{Nao Goto}
\affiliation{Department of Physical Science, Graduate School of Science, Osaka Prefecture University, 1-1 Gakuen-cho, Naka-ku, Sakai, Osaka 599-8531, Japan}

\author[0000-0001-8901-7287]{Naslim Neelamkodan}
\affiliation{Department of physics, United Arab Emirates University, Al-Ain, 15551, UAE}

\author[0000-0002-7759-0585]{Tony Wong}
\affiliation{Department of Astronomy, University of Illinois, Urbana, IL 61801, USA}

\author{Marta Sewi{\l}o}
\affiliation{CRESST II and Exoplanets and Stellar Astrophysics Laboratory, NASA Goddard Space Flight Center, Greenbelt, MD 20771, USA}
\affiliation{Department of Astronomy, University of Maryland, College Park, MD 20742, USA}

\author{Hajime Fukushima}
\affiliation{Center for Computational Sciences, University of Tsukuba, Ten-nodai, 1-1-1 Tsukuba, Ibaraki 305-8577, Japan}

\author[0000-0002-4124-797X]{Tatsuya Takekoshi}
\affiliation{Kitami Institute of Technology, 165 Koen-cho, Kitami,
Hokkaido 090-8507, Japan}

\author{Kazuyuki Muraoka}
\affiliation{Department of Physical Science, Graduate School of Science, Osaka Prefecture University, 1-1 Gakuen-cho, Naka-ku, Sakai, Osaka 599-8531, Japan}

\author{Akiko Kawamura}
\affiliation{National Astronomical Observatory of Japan, National Institutes of Natural Science, 2-21-1 Osawa, Mitaka, Tokyo 181-8588, Japan}

\author{Kengo Tachihara}
\affiliation{Department of Physics, Nagoya University, Chikusa-ku, Nagoya 464-8602, Japan}

\author{Yasuo Fukui}
\affiliation{Department of Physics, Nagoya University, Chikusa-ku, Nagoya 464-8602, Japan}

\author[0000-0001-7826-3837]{Toshikazu Onishi}
\affiliation{Department of Physical Science, Graduate School of Science, Osaka Prefecture University, 1-1 Gakuen-cho, Naka-ku, Sakai, Osaka 599-8531, Japan}

\begin{abstract}

We have analyzed the data from a large-scale CO survey toward the northern region of the Small Magellanic Cloud (SMC) obtained with the Atacama Compact Array (ACA) stand-alone mode of ALMA. The primary aim of this study is to comprehensively understand the behavior of CO as an H$_2$ tracer in a low-metallicity environment ($Z\sim0.2\,Z_{\odot}$). The total number of mosaic fields is $\sim$8000, which results in a field coverage of 0.26\,degree$^{2}$ ($\sim$2.9 $\times$10$^{5}$\,pc$^2$), corresponding to $\sim$10\% area of the galaxy. 
The sensitive $\sim$2\,pc resolution observations reveal the detailed structure of the molecular clouds previously detected in the single-dish NANTEN survey. 
We have detected a number of compact CO clouds within lower H$_2$ column density ($\sim$10$^{20}$\,cm$^{-2}$) regions whose angular scale is similar to the ACA beam size. Most of the clouds in this survey also show peak brightness temperature as low as $<$1\,K, which for optically thick CO emission implies an emission size much smaller than the beam size, leading to beam dilution. The comparison between an available estimation of the total molecular material traced by thermal dust emission and the present CO survey demonstrates that more than $\sim$90\% H$_2$ gas cannot be traced by the low-$J$ CO emission. Our processed data cubes and 2-D images are publicly available. 

\end{abstract}

\keywords{stars: formation  --- ISM: clouds--- ISM:  ---  galaxies: Local Group}

\section{Introduction} \label{sec:intro}

During the lifecycle of the interstellar medium, galaxy evolution is characterized by increasing abundance of heavy elements that are heavier than H and He, i.e., metallicity, as a consequence of star formation. The metallicity has a large impact on the cooling/heating process from interstellar gas, which results in regulating star formation therein. Observational studies toward star-forming regions in the solar neighborhood alone cannot explore a wide range of the parameter space, and extragalactic star-forming regions are thus vital targets, especially to investigate low-metallicity environments. 

The Large Maggelanic Cloud ($D\sim$50\,kpc, \citealt{de14}) and the Small Magellanic Cloud ($D\sim$62\,kpc, \citealt{Graczyk20}) are the nearest galaxies to investigate the star formation in such an environment. Their proximity enables us to obtain high spatial resolution views resolving individual star-forming regions, such as molecular cloud cores and individual protostar or protostellar systems. The metallicity of the SMC is $Z$\,$\sim$0.2$\,Z_{\odot}$ \citep{Russell92, Rolleston99, Pagel03}, which is a sufficiently sub-solar regime and lower than that of the LMC ($Z\sim$0.5\,$Z_{\odot}$, \citealt{Rolleston02}). The low-metal condition is approximate to the most active phase of star formation during cosmic history \citep{Pei99}.
It has therefore been the target of a large number of comprehensive surveys in multiple wavelengths aiming at a better understanding of the behavior of interstellar gas and young stellar objects (YSOs). The previous millimeter observations with small aperture telescopes, CfA 1.2\,m and NANTEN, revealed the overall distribution of the high-density molecular gas traced by CO \citep{Rubio91,Mizuno01} in the SMC at spatial resolutions of 150 or 45\,pc.
Spitzer/Herschel observations provided a high-dynamic-range view of cold/warm gas \citep{Bolatto07,Bolatto11,Gordon11,Gordon14,Jameson16} as well as the young stellar population \citep[e.g.,][]{Oliveira13, Sewilo13, Ruffle15}. Ground-based observations detected millimeter/submillimeter dust continuum emission from not only bright CO clouds \citep{Bot10, Hony15, Takekoshi17, Takekoshi18} but also fainter clouds, which are likely to be cold molecular material that was not accessible with the previous CO surveys \citep{Takekoshi17}.
The newly developed Australian Square Kilometre Array Pathfinder (ASKAP) interferometer has been updating the atomic hydrogen (H$\;${\sc i}) view in the SMC. The pilot study by \cite{McClure-Griffiths18} reported outflowing gas, and the excellent data set is also useful to search for some indications of the last interaction between this galaxy and the LMC based on the detailed velocity field \citep{Murray19}.

Among these surveys, the angular resolution of CO surveys was especially not high enough to investigate the detailed structures which are directly related to the star formation activities therein. Larger aperture single-dish telescopes performed $\sim$10\,pc resolution observations in CO but covered only small fields in the galaxy \citep{Rubio93,Bolatto03,Muller10}. In addition to the coarse resolution, sensitivity, and field coverage of the previous studies, CO observations intrinsically have a particular problem as a molecular gas tracer, especially in a low-metallicity environment. 
Theoretical modeling demonstrates that UV photons penetrate more deeply into dust-poor molecular clouds, and thus a large portion of molecular clouds become $``$CO dark$"$ \citep[e.g.,][]{Glover12,Fukushima20,Bisbas21}. 
In such an extreme condition, some theoretical works \citep{Glover16} and observations of distant galaxies claimed that [C$\;${\sc i}] more efficiently probes H$_2$ clouds than CO \citep{Papadopoulos04,Alaghband13}. However, \cite{Okada15, Okada19} reported that both CO and [C$\;${\sc i}] emission show similar distributions in some of the molecular clouds in the LMC. In any case, observation efficiencies with current instruments at the high-frequency (492\,GHz for [C$\;${\sc i}]) band is not high enough to obtain comprehensive maps across galaxies in the Local Group. According to \cite{Requena16}, [C$\;${\sc ii}] emission does not have significant contributions from atomic and ionized hydrogen gas and works as a better tracer of the total H$_2$ mass/column density toward their observed regions in the SMC. Nevertheless, faint CO emission enables us to trace H$\;${\sc i}--H$_2$ transition layers and overall H$_2$ structures seen in [C$\;${\sc ii}] \citep{Jameson18}. The low/mid-$J$ CO transitions in lower frequency bands are suitable for faster observations covering a large field thanks to the atmospheric condition and the larger beam size of the telescopes. 
Thus, despite its limitations, low-$J$ CO mapping remains the clearest guide to the distribution of molecular gas if sufficient sensitivity and resolution can be achieved.

Recent ALMA CO and its isotopologue observations have been revealing detailed molecular cloud distributions in the Local Group galaxies, the LMC and M33 \citep[e.g.,][]{Indebetouw13, Wong17, Naslim18, Tokuda20a, Muraoka20, Kondo21}. These studies demonstrated that $\lesssim$ a few pc resolution observations are powerful to distinguish the different properties among the observed Giant Molecular Clouds (GMCs) depending on their evolutionary stages \citep{Sawada18, Wong19} and to find indications of cloud interactions initiating high-mass star formation \citep{Fukui15a, Fukui19, Saigo17, Tokuda19, Sano21}. Although the total number of ALMA CO studies in the SMC is limited, recent observations have been revealing the density structure of molecular clouds \citep{Muraoka17, Jameson18, Naslim21}.

To reveal CO distributions in the SMC comprehensively, we need high-angular resolution data with wide spatial coverage. Use of the Atacama Compact Array (ACA, a.k.a. Morita Array) stand-alone mode is an ideal option to perform an unbiased large-scale survey as demonstrated by a dense core study in Taurus \citep{Tokuda20b}. Some CO survey programs covering a square degree scale are also ongoing in the SMC (e.g., \#2018.1.01115.S, Jameson et al. in prep). At the CO($J$ = 2--1) frequency (230\,GHz), the ACA observations provide a beam size of $\sim$6\arcsec corresponding to $\sim$2\,pc at the distance of $\sim$62\,kpc \citep{Graczyk20}. In this study, we investigated a Northern region in the SMC. One of the advantages of this region is that the velocity components in H$\;${\sc i} are relatively simple compared to the more CO bright South-West (SW) region \citep[c.f.,][]{Stanimirovic99,McClure-Griffiths18}. On the other hand, the Northern region contains several H$\;${\sc ii} regions, including the brightest one (N66, \citealt{Massey89}) in the SMC, and a few well-studied supernova remnants (SNRs), SNR~B0102-72.3 \citep[e.g.,][]{Badenes10, Maggi19} and SNR~B0057-72.2 \citep[e.g.,][]{Ye91, Naze02}. The presence of various evolutionary stages of the interstellar medium allows us to explore the behavior of CO in various environments of this low-metallicity galaxy.

We present ALMA-ACA archival CO data toward the Northern region in this paper. The data set is quite large, and thus we mostly focus on an overall description of the CO emission properties. Sect.~\ref{sec:obs} describes the observations and data reduction, and then we show the fundamental CO and continuum maps in Sect.~\ref{sec:results}. Sect.~\ref{sec:dis} gives some early analysis characterizing the CO clouds and discusses their properties.

\section{observations and data reduction} \label{sec:obs}
\begin{figure}[b]
\centering
\includegraphics[width=110mm]{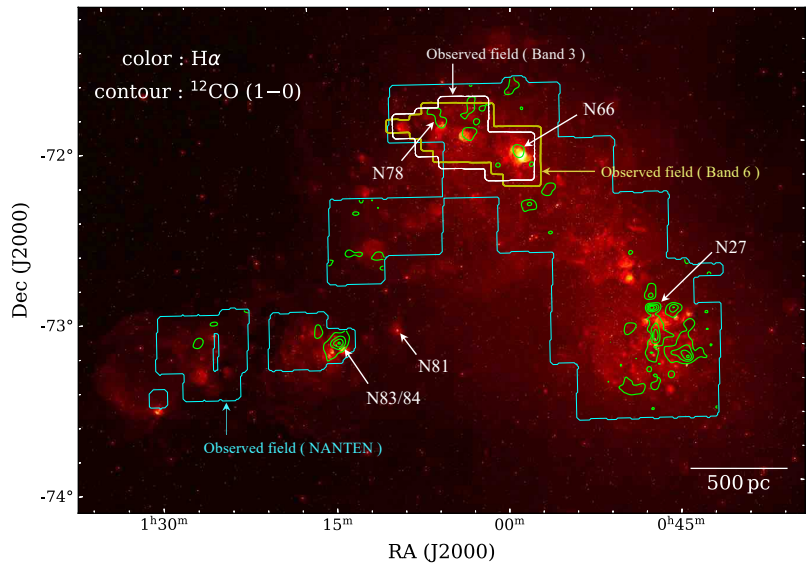}
\caption{Observed fields with the ACA stand-alone mode on an H$\alpha$ map \citep{Smith99} in the SMC. The yellow and white lines show the regions of Band~6 and Band~3 observations, respectively. The cyan lines denote the observed field boundaries with the NANTEN survey and green contours represent the CO($J$ = 1--0) map (\citealt{Mizuno01}; see also \citealt{Fukui10}). The lowest contour level and contour steps are 0.3\,K\,km\,s$^{-1}$ and 0.6\,K\,km\,s$^{-1}$, respectively. Note that the NANTEN field coverage was extended from the publication of the first survey paper (the NANTEN team, Fukui et al. private communication).
\label{fig:obs_map}}
\end{figure}

We retrieved the data from the ALMA Observatory Project 2017.A.00054.S, one of the six filler programs\footnote{https://almascience.nao.ac.jp/news/alma-announces-aca-observatory-filler-programs-for-cycle-6} for the ACA stand-alone observations in Cycle 5/6. The main target lines were CO($J$ = 2--1) in Band~6 ($\sim$230\,GHz) and CO($J$ = 1--0) in Band~3 ($\sim$115\,GHz). The total number of mosaic fields for the Band~6 and Band~3 programs were 7749 and 1939, respectively, with the ACA 7\,m array stand-alone mode. The resultant field coverage in both bands is $\sim$0.26\,degree$^{2}$, which is one of the largest observing fields among ALMA studies in Local Group galaxies. Figure~\ref{fig:obs_map} represents the observed field in Band~6+3 on the entire H$\alpha$ map in the SMC. The frequency resolution and the total bandwidth in Band~6 were 122\,kHz and 250\,MHz, respectively. For Band~3, the total bandwidth is the same, while the frequency resolution is two times better than that in Band~6 to match the velocity resolution of both bands at $\sim$0.16\,km\,s$^{-1}$. The central frequencies of the individual continuum setting were [112.5, 102.4, 100.2] GHz for Band~3, and [229.0, 215.3, 213.3] GHz for Band~6 with a 2\,GHz bandwidth in each window. The aggregate bandwidth for the continuum data integrating all of the available spectral windows was 6\,GHz.
The TP (Total Power) array observations in Band~6 were performed to supplement spatially extended emission with the same frequency setting. The beam size and sensitivity of the TP  array observations are 30\farcs6 and $\sim$0.04\,K at the native velocity resolution.
 
We used the Common Astronomy Software Application (CASA) package \citep{McMullin07} version 5.4.0 in the data reduction. We did not change the calibration scheme provided by the ALMA observatory while we performed the imaging process after concatenating all visibility files, which are separately provided in the archive. 
Although the beam shapes, i.e., sizes of the major and minor axes, of the individual observing tiles differ slightly, the aspect ratio distribution ranges 1.1-1.9 with a median value of 1.1. The final beam shape of the connected data was automatically determined by the \texttt{tclean} procedure using {\it restoringbeam} = $``$common$"$ with the natural weighting. The \texttt{multi-scale} deconvolver \citep{Kepley20} was used to recover extended emission as much as possible. The imaging grid of Band~6 and Band~3 were set to have square pixels of 2\farcs0 and 4\farcs0, respectively, and the scales of the \texttt{multi-scale} were 0, 3, and 9 pixels, which roughly corresponds to 0, 1, and 3 times of the beam size. With the \texttt{auto-multithresh} technique, we select the emission mask in the dirty and residual image. The parameters were follows; {\it sidelobethreshold} =1.25, {\it noisethreshold} =3.0, {\it lownoisethreshold} =1.5, {\it smoothfactor} = 1.2, and {\it minbeamfrac} =0.05. We continued the deconvolution process until the peak intensity of the residual image reached the $\sim$1$\sigma$ noise level. We made continuum image in the same manner except for the emission mask selection scheme of the \texttt{tclean}. We manually selected the emission mask in the continuum images because we could find only one or two significant sources (see Sect.~\ref{result:cont} and Figures~\ref{fig:Band6noise} and \ref{fig:Band3noise} in Appendix~\ref{Ap:RMScont}).

To estimate the missing flux of the Band~6 7\,m array data in CO($J$ = 2--1), we made a spatially smoothed 7\,m array data cube and then compared the flux with the TP array alone. The resultant missing flux varies from region to region (see Figure~\ref{fig:7mTPratio} in Appendix~\ref{Ap:RMScont}), but more than half of the total flux is captured by the 7\,m array alone, whose maximum recovered scale is $\sim$30$\arcsec$, corresponding to $\sim$9\,pc. Nevertheless, we combined the 7\,m and TP array data using the \texttt{feather} task to compensate for the extended emission. We used the combined CO($J$ = 2--1) data throughout this study. Table~\ref{table:data} summarizes the final beam size and the typical r.m.s sensitivity in each data. The angular resolution and sensitivity of the CO($J$ = 2--1) data are better than the 1--0 data, and we mainly present the former one throughout this paper. The 1.3\,mm and 2.6\,mm continuum data are used to identify dense materials around massive protostellar sources (Sect.~\ref{result:cont} and Appendix~\ref{Ap:RMScont}), and the CO($J$ = 1--0) maps are mainly presented in Appendix~\ref{Ap:COratio}.

To minimize the noise contribution, we made a moment-masked cube data \citep[e.g.,][]{Dame11} when we obtain moment~0,1 and 2 maps (see Sect~\ref{R:CO}). Based on a velocity/spatially smoothed data cube, we determined the emission-free pixels, which are less than $\sim$3$\sigma$ level, and set them to zero value. An additional criterion for creating emission masks is whether the CO emitting regions more than $\sim$3$\sigma$ are continuously connected to each other over 20 voxels, and thus very tiny features whose size is smaller than the resolution elements are ignored. However, The fraction of the real emissions that were not selected as positive masks is not large because the masked moment maps fairly reproduce the overall distributions of the peak temperature map, which is made of the unmasked cube data (see Figures~\ref{fig:12COii} and \ref{fig:moms} in Sect.~\ref{R:CO}.)

\begin{table}[ht]
\centering
\caption{Summary of the processed data quality}
\scalebox{1}{
\begin{tabular}{lcccc} \hline \hline
Data & Beam size & Velocity resolution (km\,s$^{-1}$) & r.m.s (K) & r.m.s (Jy\,beam$^{-1}$)\\
\hline
CO($J$ = 2--1)   & 6\farcs9$\times$6\farcs6 & 0.5 & $\sim$0.06 & $\sim$9$\times\,$10$^{-2}$ \\
CO($J$ = 1--0)   & 14\farcs9$\times$11\farcs3 & 0.5 & $\sim$0.22 & $\sim$4$\times\,$10$^{-1}$ \\ 
1.3\,mm continuum & 7\farcs2$\times$6\farcs7 & $\cdots$ & $\cdots$ & $\sim$2$\times\,$10$^{-3}$ \\ 
2.6\,mm continuum & 15\farcs8$\times$12\farcs6 & $\cdots$ & $\cdots$ & $\sim$2$\times\,$10$^{-3}$ \\ 
\hline
\end{tabular}}
\label{table:data}
\end{table}

The final reduced fits images, CO cubes and continuum 2-D maps, are available at \dataset[10.5281/zenodo.4628967]{https://doi.org/10.5281/zenodo.4628967}.

\section{Results} \label{sec:results}
\subsection{Overall Distribution of CO($J$ = 2--1) clouds in the SMC North region}\label{R:CO}
 
\begin{figure}[htbp]
\includegraphics[width=180mm]{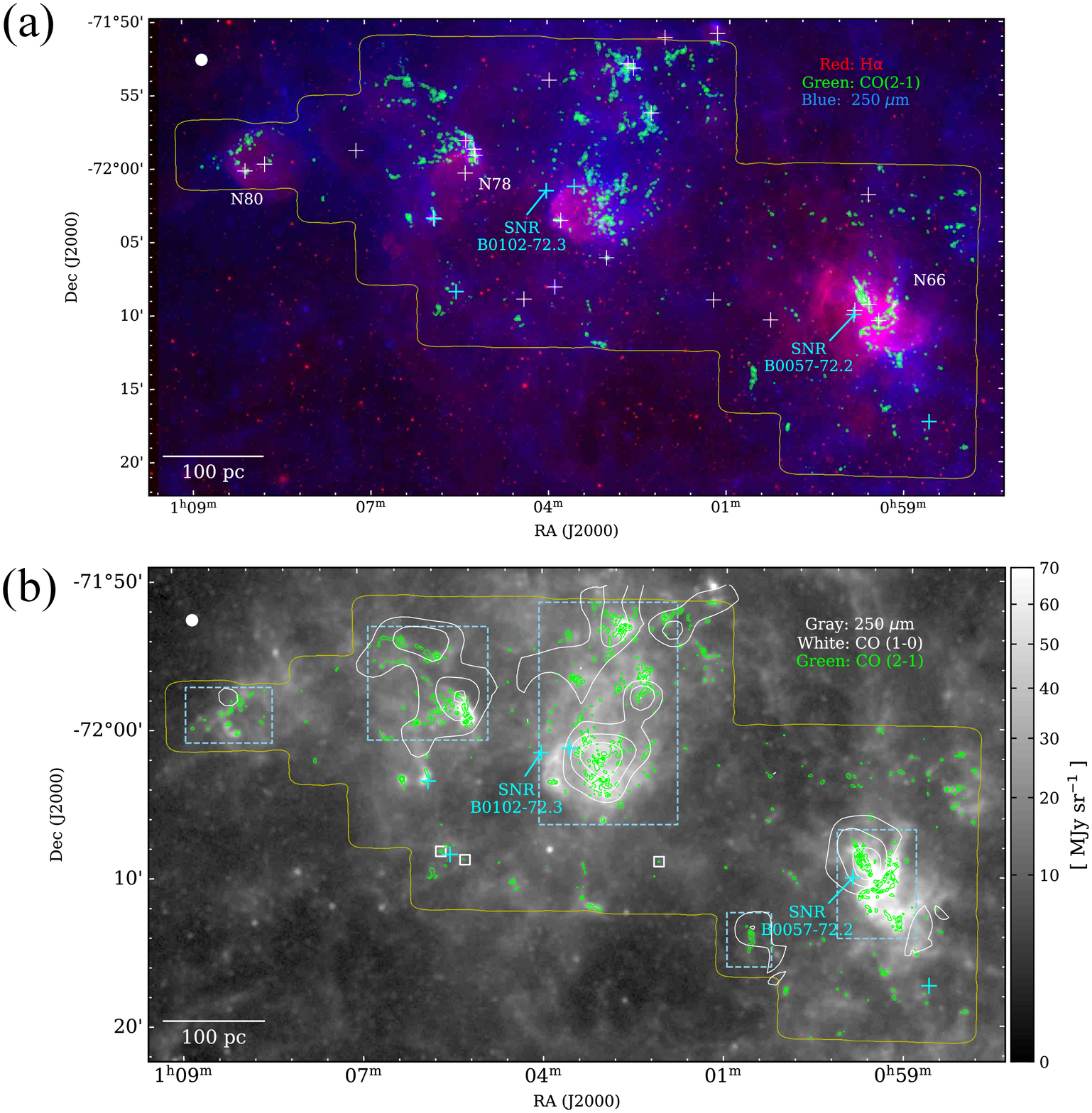}
\caption{The integrated intensity (moment~0) map of CO($J$ = 2--1) toward the SMC North region obtained by the ACA stand-alone mode. (a) The three-color composite image combining the ACA integrated intensity (moment 0) CO($J$ = 2--1) (green) H$\alpha$ (red) \citep{Smith99}, and Herschel/SPIRE 250$\mu$m \citep{Gordon14} maps. The green contours correspond to the CO($J$ = 2--1) emission with contour levels of [1, 6, 11, 16] K\,km\,s$^{-1}$. The yellow line denotes the field coverage of the ACA observations. The circle in the upper-left corner shows the size of the single mosaicing field with the ACA. The white and cyan crosses represent positions of H$\;${\sc ii} regions \citep{Henize56} and SNRs \citep{Maggi19}, respectively. 
(b) Same as (a) but for the Herschel/SPIRE 250$\mu$m map in gray-scale. The white contours show the NANTEN CO($J$ = 1--0) map. The area enclosed by the cyan dashed rectangles roughly represent the regions with the known CO clouds detected by the NANTEN survey (see also the text in Sect.~\ref{dis:ACAclouds}). The three small squares represent the examples of compact CO clouds shown in Figure~\ref{fig:ACAclouds} (see Sect.~\ref{dis:ACAclouds}).}
\label{fig:12COii}
\end{figure}

Figure~\ref{fig:12COii} shows the integrated intensity CO($J$ = 2--1) map toward the SMC North region. For the first time, the ACA observations have revealed the presence/absence of CO clouds and their distributions in this particular part of the galaxy with a high spatial dynamic range whose size scale is from $\sim$2\,pc to more than $\sim$1\,kpc.
As shown in panel (a,b), the relatively bright/large CO clouds are distributed in and around the H$\;${\sc ii} regions (LHA 115-N N66 (N66), LHA 115-N N78 (N78), and LHA 115-N N80 (N80), \citealt{Henize56}) and SNRs (B0102-72.3, B0057-72.2).  The locations of these clouds on the Herschel map (Panel (b)) are close to the local peaks of the continuum emission.  Although these clouds are mostly detected by the previous NANTEN survey \citep{Mizuno01}, we successfully detected much smaller/faint clouds, whose size scale is a few parsecs across the observed field, around the large clouds. These clouds are real emission instead of noise-like features (see also some examples shown in Sect.~\ref{dis:ACAclouds}). They do not necessarily correlate with the local peak of the Herschel dust emission and the minimum intensity of 250\,$\mu$m emission with CO emission is $\sim$5\,MJy\,sr$^{-1}$. In Sect.~\ref{dis:COprop}, we further discuss how CO emission traces molecular clouds in this galaxy.

\begin{figure}[htbp]
\includegraphics[width=170mm]{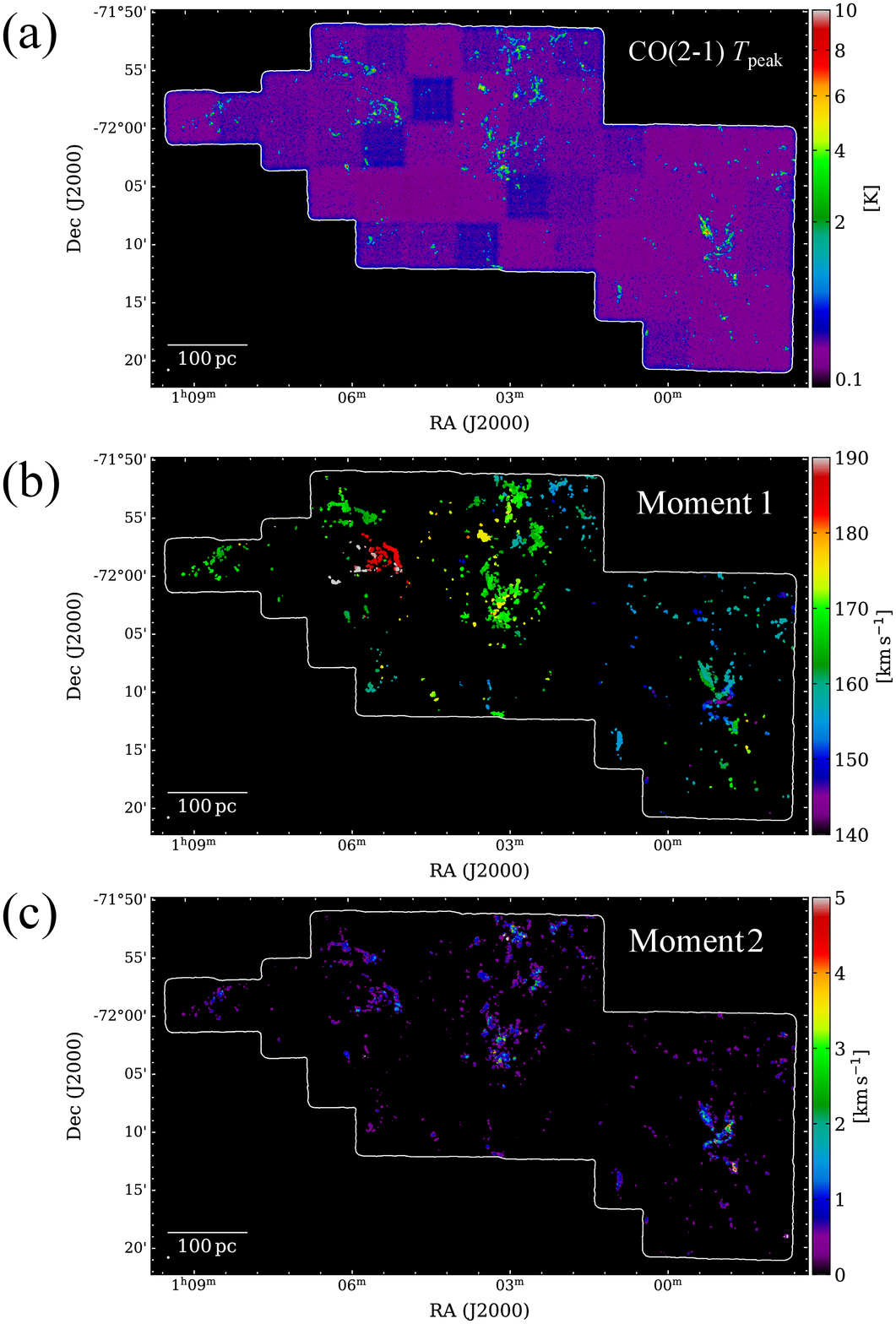}
\caption{Panel (a), (b), and (c) show the CO($J$ = 2--1) peak brightness temperature, centroid velocity (moment~1), and velocity dispersion (moment~2) maps, respectively.  
\label{fig:moms}}
\end{figure}

Figure~\ref{fig:moms} (a) represents the CO peak brightness temperature ($T_{\rm peak}$) map. 
The typical temperature is a few kelvins throughout the observed field, which is remarkably weaker than 10\,K, although similar spatial resolution studies in the MW (Milky Way) and M33 show more bright emission greater than 10K. One of the possible interpretations to explain the faint nature of CO emission is the beam dilution effect. The CO clouds are too small to spatially resolve even with the $\sim$2\,pc resolution (see more detailed discussion in Sect.~\ref{dis:COprop}). The moment~1 map in panel (b) shows the velocity distribution in CO. The dominant velocity components in this region are $\sim$160--170\,km\,s$^{-1}$. Some clouds show velocity gradient across their minor/major elongations; the others show more complex velocity fields. We found a remarkable difference: the N78 region is $\sim$10--20\,km\,s$^{-1}$ redshifted compared to the average velocity. Panel (b) illustrates the moment~2 map with a displayed range of 0--5\,km\,s$^{-1}$. The large clouds (see the previous paragraph) tend to show larger velocity dispersion, $\sim$2--3\,km\,s$^{-1}$, around their CO peak in the moment~1 or $T_{\rm peak}$ maps.   

\begin{figure}[htbp]
\includegraphics[width=180mm]{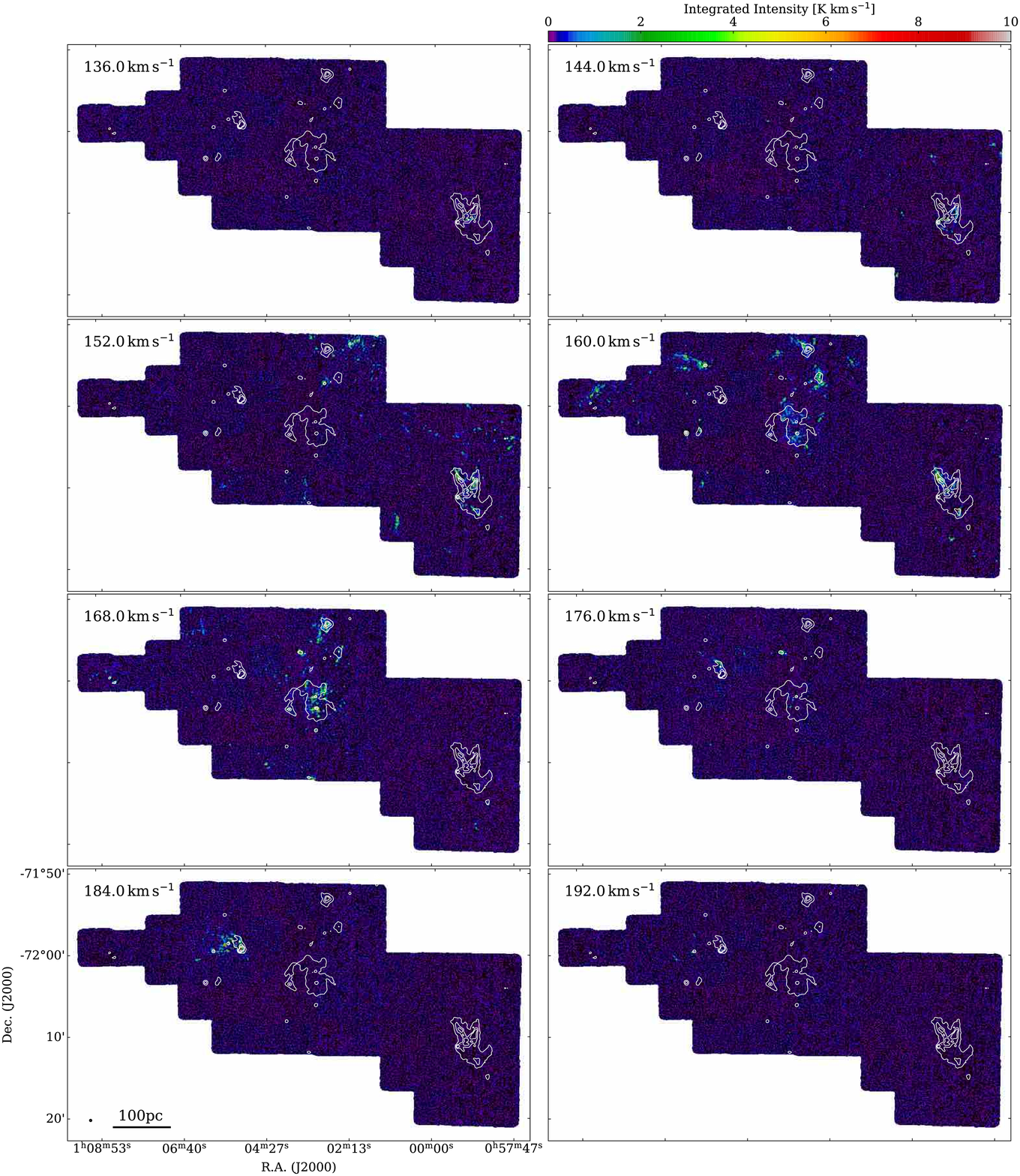}
\caption{Velocity-channel maps toward the SMC North in CO($J$ = 2--1). The lowest velocity is shown in the upper left corner of each panel. The beam size of the ACA CO observations is shown by the black ellipse in the lower-left corner of the lower-left panel. The white contours show the Herschel/SPIRE 250\,$\mu$m map with the contour levels of [40, 80, 120] MJy\,sr$^{-1}$. 
\label{fig:chanmap}}
\end{figure}

Figure~\ref{fig:chanmap} illustrates the velocity-channel map with the Herschel 250$\mu$m contours as reference positions. Most of the bright dust emitting regions are visible in CO (see also Figure~\ref{fig:12COii}). 
The representative velocity of the N66, SNR~B0102-72.3, N78, and N80 regions are 160, 168, 184, 160\,km\,s$^{-1}$, respectively. 

\subsection{Millimeter continuum sources in the N78 region}\label{result:cont}

\begin{figure}[htbp]
\includegraphics[width=180mm]{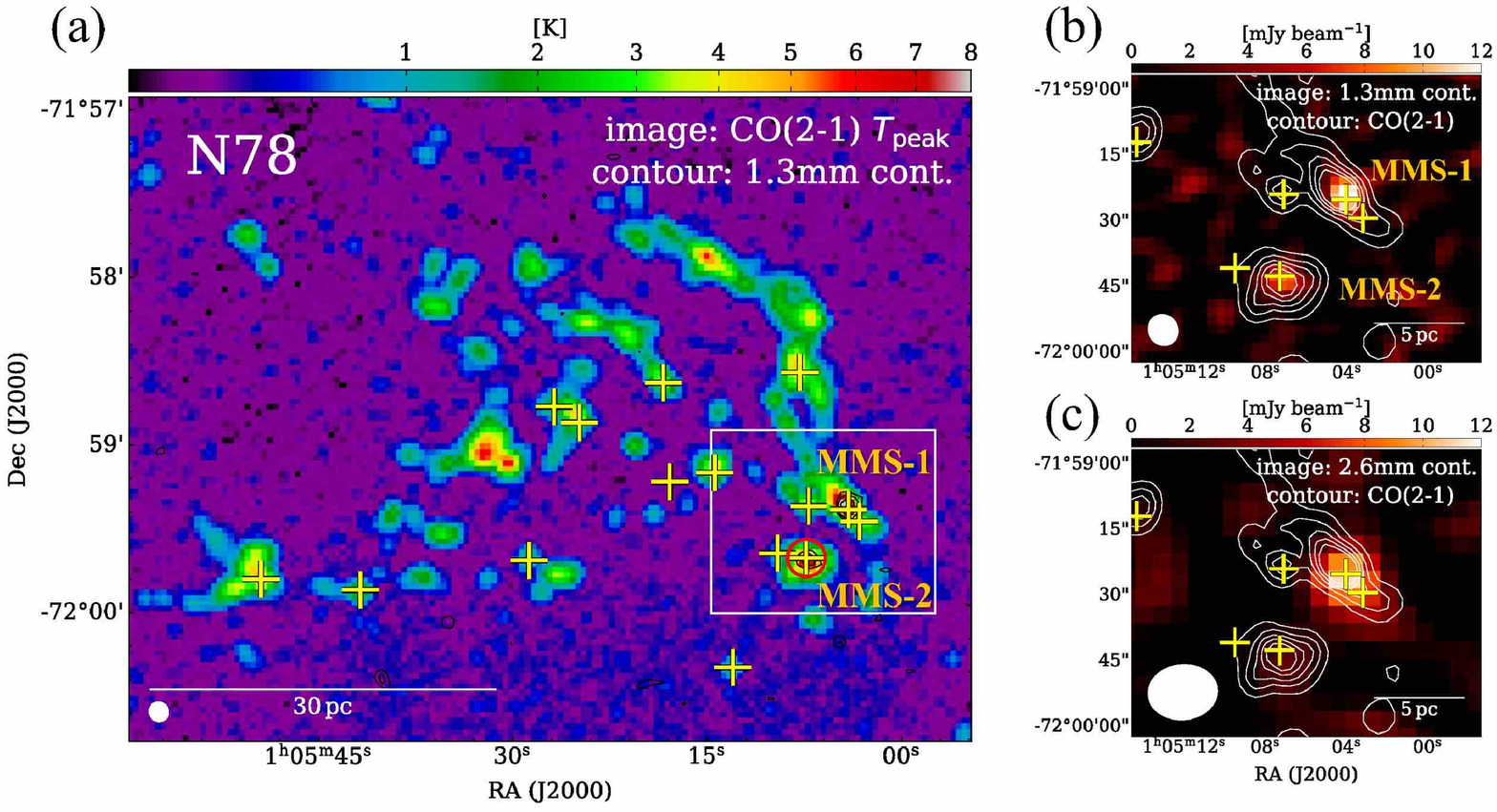}
\caption{(a) An enlarged view of the CO($J$ = 2--1) peak brightness temperature ($T_{\rm peak}$) map around the N78 region. The black contours show the 1.3\,mm continuum image with the contour levels of [5,7,10] mJy\,beam$^{-1}$. The white rectangle corresponds to the visualized area in panel (b,c). The white ellipse in the lower-left corner shows the beam size of CO($J$ = 2--1). The yellow crosses denote the positions of the YSO candidates \citep{Sewilo13}. The red circle is a confirmed YSO by the spectroscopic observation \citep{Oliveira13}.
(b) Color-scale and contours represent the 1.3\,mm continuum and CO($J$ = 2--1) $T_{\rm peak}$ maps, respectively. The lowest CO($J$ = 2--1) contour level and contour step are 1.0\,K, and 1.0\,K, respectively. The white ellipse in the lower-left corner gives the beam size of the continuum image. (c) Same as panel (b) but for the Band~3 (2.6\,mm) continuum data.
\label{fig:N78_cont}}
\end{figure}

We detected significant ($>$3$\sigma$) continuum emission only around CO intense positions in the N78 region. Figure~\ref{fig:N78_cont} shows a zoomed-in view of the CO($J$ = 2--1) and continuum maps. We detected at least two continuum sources in 1.3\,mm (see panel(b)): the measured fluxes of the northern (MMS-1) and southern (MMS-2) sources are $\sim$230\,mJy and $\sim$130\,mJy, respectively. MMS-1 is bright in 2.6\,mm also (panel (c)), whose flux is $\sim$160\,mJy. The spectral index derived from the two bands in the northern source is $\sim$0.5, indicating that the continuum flux in MMS-1 is likely dominated by the free-free emission rather than thermal dust emission. 

We found YSO (candidate) associations to the two millimeter sources based on the \cite{Sewilo13} catalog. The source in MMS-2 is the most luminous source ($L$ $\sim$1.5\,$\times$10$^{5}$\,$L_{\odot}$) within the ACA field. The estimated stellar mass is $\sim$28\,$M_{\odot}$. \cite{Oliveira13} confirmed that it is indeed a YSO (\#28) using their spectroscopic observations. The millimeter continuum emission is useful to probe the presence of massive/dense clumps around a massive protostar. The total 1.3\,mm flux of MMS-2 infers that the gas mass is $\sim$6\,$\times$10$^{3}$\,$M_{\odot}$ assuming the gas-to-dust ratio of 1000 (e.g., \citealt{Roman-Duval14}), dust opacity of 1\,cm$^{-2}$\,g$^{-1}$ (e.g., \citealt{Ossenkopf94}), and dust temperature of 24\,K \citep{Takekoshi18}.

\section{Discussion} \label{sec:dis}

\subsection{Mass Estimation of CO clouds and the Detection Limit of the Survey}\label{dis:Xco}
There is no unique way to estimate molecular cloud mass based on the CO observation, mostly due to the uncertainty in the CO($J$ = 1--0)-to-H$_2$ conversion ($X_{\rm CO}$) factor (see the review by \citealt{Bolatto13}) and intensity ratios of CO($J$ = 2--1) and CO($J$ = 1--0) (hereafter, $R_{2-1/1-0}$) throughout the galaxy. Nevertheless, we tentatively estimate the mass detection limit based on the survey sensitivity and resolution elements. The 1$\sigma$ noise level of the data cube of $\sim$0.06\,K and the typical line width of the CO clouds of 1.5\,km\,s$^{-1}$ make that the 1$\sigma$ noise level of the integrated intensity is $\sim$0.05\,K\,km\,s$^{-1}$. We tentatively set the detection criteria as that the CO emission region is larger than the single beam element ($\sim$2\,pc) at more than 3$\sigma$, and the detection limit in the CO luminosity itself is $\sim$1.0\,K\,km\,s$^{-1}$ pc$^{2}$. 

\cite{Bolatto03} performed CO($J$ = 2--1) and CO($J$ = 1--0) observations toward the SMC N83/N84 clouds with SEST and derived average $R_{2-1/1-0}$ as $\sim$0.9. Although the sensitivity is not high enough to derive $R_{2-1/1-0}$ at the native resolution of CO($J$ = 1--0), we obtained a similar result, $R_{2-1/1-0}$ $\sim$1 (see Appendix~\ref{Ap:COratio}). 
These values are remarkably higher than those in the MW average, $\sim$0.6 \citep{Yoda10}, and the other CO bright nearby galaxies, which are 0.6–0.8 \citep{Yajima21}. In the typical MW molecular clouds, the $J$ = 2--1 line is not fully thermalized and $R_{2-1/1-0}$ does not approach to 1 \citep{Sakamoto94, Yoda10, Nishimura15}. One possible explanation of this discrepancy is that the SMC CO observations trace a much deeper side of molecular clouds whose density is $\sim$10$^{4}$\,cm$^{-3}$ (see \citealt{Muraoka17}). In this case, densities of CO emitting region are close to or higher than the critical density of $J$ = 2--1 line, and $R_{2-1/1-0}$ becomes $\sim$1. Numerical modelings of molecular cloud adopting the SMC-like metallicity also reproduced a similar ratio, possibly due to the proposed reason \citep[c.f.,][]{Bisbas21}, i.e., tracing a higher density region than the MW by the CO observations. Sect.~\ref{dis:COprop} further explains the high-density nature of CO clouds in the observed region. In summary, our tentative use of $R_{2-1/1-0}$ as 0.9--1.0 seems to be reasonable based on the available measurements.

\cite{Mizuno01} estimated the $X_{\rm CO}$ factor in the SMC (hereafter, $X_{\rm CO}^{\rm SMC}$) as 2.5\,$\times$10$^{21}$\,cm$^{-2}$\,(K\,km\,s$^{-1}$)$^{-1}$ assuming of the virial equilibrium throughout the observed cloud. Their estimation is an order of magnitude higher than the Milky Way standard value (e.g., \citealt{Dame01, Bolatto13}), however the larger beam size ($\sim$45\,pc) likely makes a large uncertainty of the cloud radii to derive the virial mass.  
Higher angular resolution CO data with the same analysis estimated $X_{\rm CO}^{\rm SMC}$ as (4--7.5) \,$\times$10$^{20}$\,cm$^{-2}$\,(K\,km\,s$^{-1}$)$^{-1}$ \citep{Bolatto03, Muraoka17}.
The previous measurements tell us that the conversion factor is a few or several times larger than $X_{\rm CO}$ in the MW. We tentatively use $X_{\rm CO}$ of $\sim$6\,$\times$10$^{20}$\,cm$^{-2}$\,(K\,km\,s$^{-1}$)$^{-1}$ (= $\alpha_{\rm CO}$ of 13\,$M_{\odot}$\,(K\,km\,s$^{-1}$\,pc$^{2}$)$^{-1}$), which is mean value from the literature \citep{Bolatto03, Muraoka17}, and $R_{2-1/1-0}$ of 0.9 to estimate the mass of the CO clouds in this study. The uncertainty is supposed to be a factor of $\sim$2. In this case, the luminosity detection criteria of $\sim$1.0\,K\,km\,s$^{-1}$ pc$^{2}$ can be converted into $\sim$10--20\,$M_{\odot}$ as the mass detection limit. Note that the current large-data set can be used to estimate the $X_{\rm CO}$ factor as well, but the method is not decoupled from a complex cloud identification scheme. Although we partially applied cloud decomposition analysis toward some of the simple/isolated clouds in the observed field (Sect.~\ref{dis:ACAclouds}), a separate paper will perform further comprehensive analyses to obtain the full cloud catalog, including their individual physical properties and $X_{\rm CO}$ factor.

\subsection{CO cloud properties and the behavior of CO as a tracer} \label{dis:COprop}

\subsubsection{Compact CO clouds}\label{dis:ACAclouds}

The angular resolution and mass detection sensitivity of the NANTEN survey were 45\,pc and 10$^4$\,$M_{\odot}$, respectively, which allows us to identify GMCs in terms of studies in the MW-like galaxies. Similar angular resolution studies in LMC/M33 and nearby galaxies with large-aperture single dishes (e.g., \citealt{Fukui08, Tosaki11, Onodera10, Onodera12, Miura12, Corbelli17}) and interferometers \citep[e.g.,][]{Schinnerer13, Pety13, Sun18, Brunetti21} discovered more than a few hundred GMCs per galaxies. However, the CO emission in the SMC is remarkably weaker than that in the above galaxies, and thus \cite{Mizuno01} predicted a clumpy CO distribution within the NANTEN beam. The present ACA survey revealed the NANTEN identified sources are likely molecular cloud clusters rather than an individual GMC, at least in the CO traced view, which is reasonably consistent with the \cite{Mizuno01} prediction (see also the enlarged and high-resolution view of N66 in Sect.~\ref{sec:N66}). 

There are many isolated/compact clouds outside the NANTEN-detected regions (see the cyan rectangles in Figure\,\ref{fig:12COii}). After the NANTEN survey \citep{Mizuno01}, the subsequent follow-up CO studies in the SMC North region \citep{Muller10} did not observe the outside regions. Any previous single-dish observations thus could not find such compact entities. We call them $``$the compact CO clouds$"$ hereafter. The comparison between the CO and Herschel maps tells us that the submillimeter (250\,$\mu$m) continuum emission does not exceed more than 25\,MJy\,sr$^{-1}$ at the location of the compact CO clouds. This result means that the compact CO clouds are embedded at the lower hydrogen column density regions if we assume the thermal dust emission is approximately proportional to the total hydrogen material. 

We characterize the properties of the compact CO clouds. Because most of them are spatially well-separated at the lowest contour level (see Figure\,\ref{fig:12COii}), identifying the individual components is straightforward compared to more complex NANTEN-detected clouds. We applied the \texttt{astrodendro} algorithm \citep{Rosolowsky08} to the CO($J$ = 2--1) moment-masked data cube (see Sect.~\ref{sec:obs}) outside the dashed cyan rectangles in Figure~\ref{fig:12COii}(b). There are three input parameters, \texttt{min\_value}, \texttt{min\_delta}, and \texttt{min\_npix} in the algorithm. The first argument is the minimum intensity value to consider in the cube data; we set this value as 0\,K to capture weaker emission as much as possible. The second is the threshold value for entities in close proximity to each other to be regarded as independent components; our adapted value is 0.18\,K, corresponding to $\sim$3\,$\sigma$ noise level of the cube data. We set \texttt{min\_npix}, which is the minimum voxel number with significant emission, as 53 that equivalents to the number having at least a single beam element in XY space and three pixels in the velocity direction. We extracted the largest continuous structures, called $``$trunk$"$. The resultant boundaries of the identified sources are almost same as the lowest contour level on their moment~0 map (Figure~\ref{fig:12COii}). We estimated the physical quantities of the compact CO clouds, peak brightness temperature ($T_{\rm peak}$), velocity dispersion ($\sigma_{v}$), CO($J$= 2--1) luminosity ($L_{\rm CO(2-1)}$), beam-deconvolved radius ($R_{\rm deconv}$), H$_2$ column density ($N_{\rm H_2}$), total molecular mass ($M_{\rm CO}$), and average volume density ($n_{\rm H_2}$). 

Table~\ref{table:COclouds} shows the typical (median) physical parameters of the compact CO clouds. The total mass of the other CO clouds inside the cyan rectangles in Figure\,\ref{fig:12COii}, is calculated to be $\sim$3.2$\times$10$^{5}$\,$M_{\odot}$. The mass fraction of the compact CO clouds is $\sim$20\% with respect to the total mass within the observed field. Since the previous NANTEN survey with a 45\,pc resolution was not able to detect such compact emission, these high sensitivity/angular resolution observations are key to reveal a full population of CO clouds in the low-metallicity environment.
One of the notable characteristics is the low peak brightness temperature of less than 1\,K. This feature is applicable to not only the compact CO clouds but also most of the other clouds in the present observed field (see Figure~\ref{fig:moms}(a)). Suppose the CO emission is optically thick, which is basically applicable to the Galactic molecular clouds; a small beam filling factor results in the observed temperature being well below 10–20\,K, the typical temperature of molecular clouds.
In other words, the CO clouds are not fully resolved yet even with $\sim$2\,pc resolution (see the discussion in NGC~6822 by \citealt{Schruba17}), and thus the derived column density and density are lower limits.

\begin{table}[htbp]
\caption{Summary of median physical properties of the compact CO clouds in the SMC North region}
\begin{flushleft}
\vspace{-0.5cm}
\begin{tabular}{ccccccccc} \\ \hline \hline
Number & $T_{\rm peak}$ & $\sigma_{v}$ & $L_{\rm CO(2-1)}$ & $R_{\rm deconv}$ & $N_{\rm H_2}$ & $M_{\rm CO}$& $n_{\rm H_2}$ & total $M_{\rm CO}$ \\
 & [K] & [km s$^{-1}$] & [K km s$^{-1}$ pc$^{2}$] & [pc] & [10$^{21}$ cm$^{-2}$] & [$M_{\odot}$] & [10$^{2}$ cm$^{-3}$] & [10$^{5}$ $M_{\odot}$] \\
 (1) & (2) & (3) & (4) & (5) & (6) & (7) & (8) & (9) \\ \hline
 153 & 1.3 & 0.52 & 11.7 & 1.30 & 1.1 & 167 & 7 & 0.7 \\ \hline
\end{tabular}
\end{flushleft}
\vspace{-0.5cm}
\footnote[0]{
(1) Total number of the identified clouds 
(2) Peak brightness temperature 
(3) Velocity dispersion 
(4) Integrated CO(2--1) luminosity 
(5) Beam deconvolved radius, $R_{\rm deconv} = \sqrt{R_{\rm obs}^2 - R_{\rm beam}^2}$, where $R_{\rm obs}$ is the geometric mean of the effective rms sizes in major and minor axes multiplied by a factor of 1.91 as suggested by \cite{Solomon87} and $R_{\rm beam}$ is the beam size in pc
(6) Peak H$_2$ column density with the assumptions of $X_{\rm CO}^{\rm SMC}$ = 6$\times$10$^{20}$\,cm$^{-2}$\,(K\,km\,s$^{-1}$)$^{-1}$ and $R_{\rm 2-1/1-0}$ = 0.9 (see the text in Sect.~\ref{dis:Xco}) 
(7) Gas mass with the same assumptions of column~6 
(8) Average H$_2$ volume density, $3M_{\rm CO}/4\pi\mu m_{\rm H}R_{\rm deconv}^3$ where $\mu$ is the molecular weight per hydrogen (2.7) and $m_{\rm H}$ is the H atom mass 
(9) Cumulative $M_{\rm CO}$ in the observed clouds}
\label{table:COclouds}
\end{table}

\begin{figure}[htbp]
\centering
\includegraphics[width=180mm]{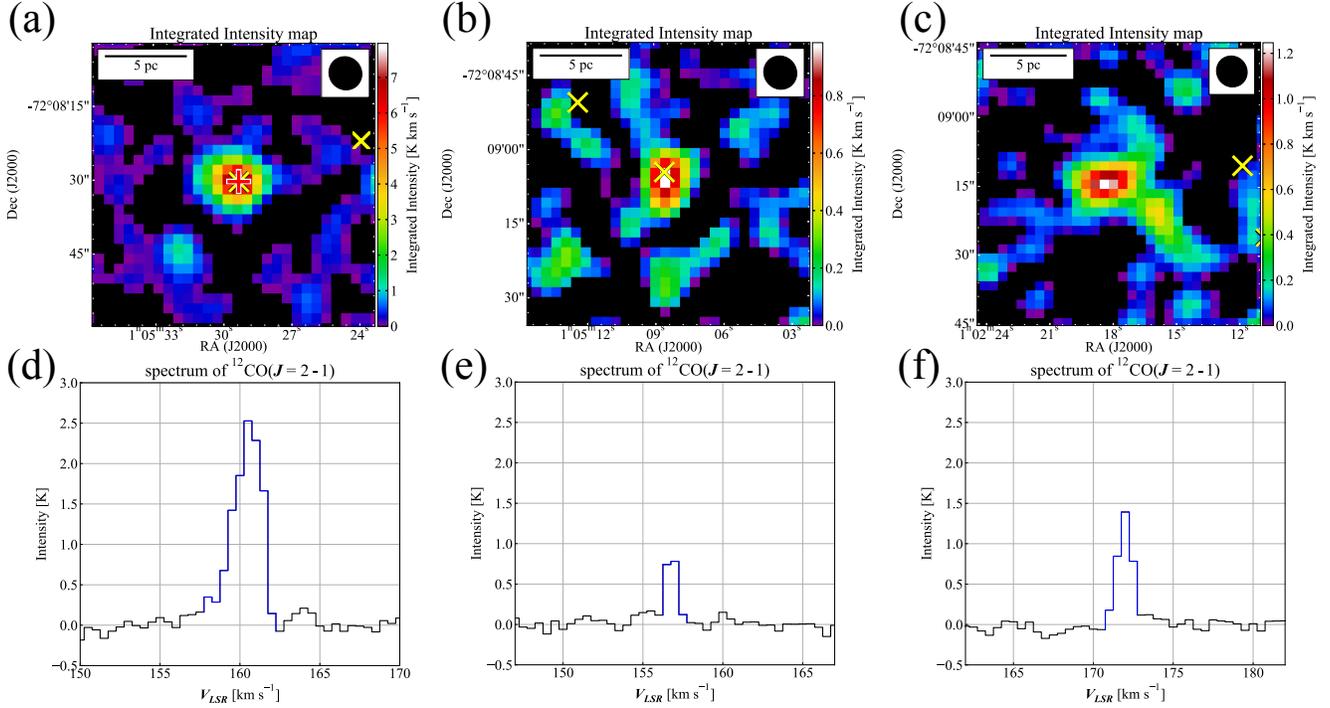}
\vspace*{0.5cm}
\caption{Enlarged-views and their spectra of three of the compact CO clouds (see also the locations as indicated in the white squares in Figure~\ref{fig:12COii}(b)). (a) The CO($J$ = 2--1) integrated intensity (moment~0) map toward one of the compact CO clouds with the YSO candidate. The emission was integrated over the velocity range from 158.0 to 162.0\,km\,s$^{-1}$ (shown in blue in panel (d)). The beam size of the ACA CO observations is shown in the black ellipse of the upper-right corner. The red and yellow crosses are the high-reliability YSO candidate (Y698) \citep{Sewilo13} and 8\,$\mu$m sources from the SAGE-SMC catalog \citep{Gordon11}, respectively. (b,c) Same as (a) but for the cloud associated with and without 8\,$\mu$m point source, respectively. The integrated velocity ranges are 156.5--157.5\,km\,s$^{-1}$ for panel (b) and 171.0--172.5\,km\,s$^{-1}$ for panel (c). (d) The CO($J$ = 2--1) spectrum extracted from a single-pixel corresponding to the CO peak on panel (a). The highlighted velocity range, 158.0--162.0\,km\,s$^{-1}$ in blue-color, is the significant emission above 3$\sigma$ noise level. (e,f) Same as (d) but for the sources in panel (b,c), respectively. }
\label{fig:ACAclouds}
\end{figure}

Figure\,\ref{fig:ACAclouds} shows the enlarged views and spectra toward three compact CO clouds. The selection is based on comparing the YSO candidates and the 8$\mu$m point source catalog, which is mainly obtained by the Spitzer survey \citep{Gordon11,Sewilo13}. The spatial extension is similar to the beam size, and the peak brightness temperature is 1--2\,K, indicating that these clouds are not spatially resolved, as discussed in this section. Although a forthcoming paper (Ohno et al. submitted) provides a further detailed comparison between the infrared sources and the CO properties, we highlight three examples in Figure~\ref{fig:ACAclouds} and discuss what they are. 

The left panels show one of the sources associated with a YSO candidate \citep{Sewilo13}. This particular YSO (Y698) candidate is one of the Stage~I sources with a luminosity of $\sim$3$\times$10$^3$\,$L_{\odot}$ and a protostellar mass of $\sim$10\,$M_{\odot}$. The YSO position fairly corresponds to the CO peak, indicating that the YSO is in an early evolutionary stage of high-mass star formation without destroying the parental cloud. In the LMC, \cite{Harada19} discovered similar compact CO clouds, which are more than $\sim$200\,pc apart from nearby GMCs, associated with massive YSO candidates. They also explained that at least one of their targets harbors a stellar cluster instead of a single O-type star. Combining our new findings in the SMC, we demonstrated that such star-forming compact CO clouds do exist in both of the Magellanic Clouds. Some theoretical studies suggest that gas accretion of CO dark H$_2$ envelope or atomic hydrogen are not negligible as mass supply onto protostars in metal-poor conditions \citep{Krumholz12,Fukushima20}. Additional high-resolution molecular line and H$\;${\sc i} studies are helpful to advance our understanding of the detailed gas accretion process onto the protostars.

We confirmed that the CO peak and the Spitzer 8\,$\mu$m source (see the catalog in \citealt{Gordon11}) has a good spatial correspondence in the middle panel sources (Figure\,\ref{fig:ACAclouds} (c,d)). Although \cite{Sewilo13} did not catalog this source as a YSO candidate, the good correlation between the source position and molecular gas indicates that the infrared source is not physically unrelated objects, such as external galaxies and evolved stars. Further infrared wavelength analysis is needed to characterize the properties of the central Spitzer source. The right panels show one of the sources without any infrared sources observed by the Spitzer, indicating that the star formation activity is quiescent compared to the other two sources. We cannot prove the CO clouds are purely in a starless phase down to solar and sub-star mass stars. Nevertheless, these infrared-free sources in the SMC may be vital targets to investigate the initial condition of star formation in the metal-poor galaxy.   

\subsubsection{Enlarged View toward the N66 region and comparison with the MW Orion molecular clouds}\label{sec:N66}
\begin{figure}[htbp]
\includegraphics[width=180mm]{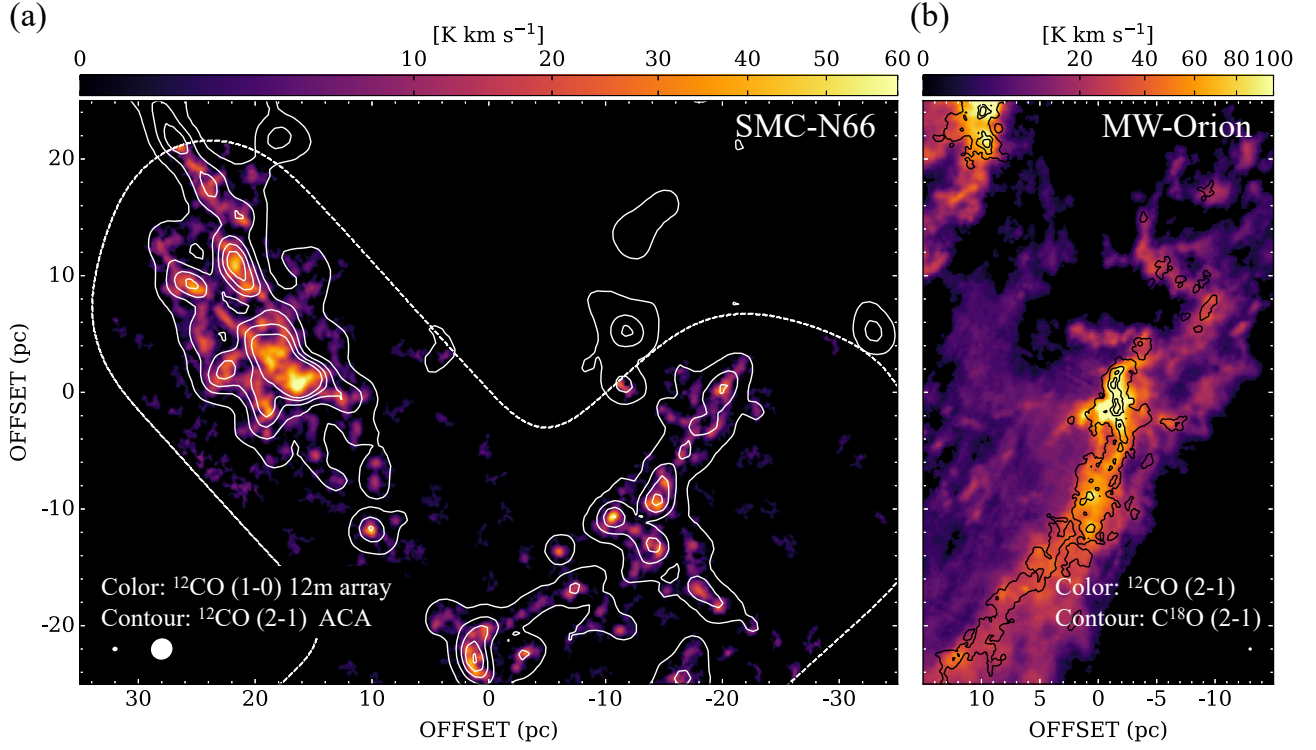}
\caption{CO distributions of the SMC-N66 region and the Orion molecular cloud in the MW at the same linear scale. (a) The color-scale and contours show the velocity-integrated intensity of CO($J$ = 1--0) \citep{Naslim21} and CO($J$ = 2--1), respectively. The dotted lines denote the observed region in CO($J$ = 1--0). The lowest CO($J$ = 2--1) contour level and contour step are 1\,K\,km\,s$^{-1}$ and 4\,K\,km\,s$^{-1}$, respectively. The small and large ellipses at the lower left corner are the beam sizes of CO(2--1) ACA and CO(1--0) 12\,m array data, respectively.
(b) The color-scale and contours show the velocity-integrated intensity of CO($J$ = 2--1) and C$^{18}$O($J$ = 2--1), respectively \citep{Nishimura15} obtained with the Osaka 1.85\,m telescope \citep{Onishi13}. The lowest C$^{18}$O($J$ = 2--1) contour level and contour step are 1\,K\,km\,s$^{-1}$ and 3\,K\,km\,s$^{-1}$, respectively.
\label{fig:N66Orion}}
\end{figure}

Figure~\ref{fig:N66Orion} presents the comparison between the CO intense cloud, the N66 region, whose moment~0 and peak temperature are the highest in this survey (see Figures\,\ref{fig:12COii} and \ref{fig:moms}(a)), and the Orion molecular clouds in the MW at the same linear scale. The displayed area in panel (a) is located in/around the H$\;${\sc ii} region N66, the largest and most luminous one in the SMC, and hosts nearly 33 OB stars \citep{Massey89,Walborn00,Evans06}. The Orion clouds are one of the reasonable targets to show the gas distribution as the typical and well-studied high-mass star-forming region in the MW, although its star formation activity \citep[e.g.,][]{Hill97} is not extreme compared to N66. We supplement the ACA map with an alternative ALMA CO($J$ = 1--0) data of the N66 region obtained by the ALMA 12\,m array \citep{Naslim21} to show the higher resolution view. Note that we confirmed the 12\,m array data has no significant missing flux based on the comparison between the single-dish SEST spectrum \citep{Rubio96} and a smoothed ALMA data at the same angular resolution of 43$\arcsec$.

The N66 region does not show remarkable extended emission in CO, whose size is more than a few tens of parsecs, as seen in Orion. The higher resolution 12\,m data shows elongated structures. The spatial extent of CO in N66 is similar to those of C$^{18}$O in Orion. Although additional work is required to precisely estimate the actual density traced by the CO observations, CO in the SMC may preferentially trace the innermost part of molecular clouds, whose density is similar to that traced by the C$^{18}$O observation in the MW \citep[e.g.,][]{Onishi96}. In fact, ALMA observations by \cite{Muraoka17} reported that the CO observations in the SMC-N83 region trace $\sim$10$^{4}$\,cm$^{-3}$ density gas based on their non-LTE (local thermodynamical equilibrium) analysis using multi-line CO/$^{13}$CO data (see also a lower resolution study by \citealt{Requena16} in N66). 
This nature is consistent with the numerical prediction that CO dramatically becomes abundant  more than an H$_2$ volume density of $\sim$10$^{4}$\,cm$^{-3}$ \citep{Glover11,Glover12}. 

\subsubsection{Column density estimation based on the CO and thermal dust emission}

\begin{figure}[htbp]
\includegraphics[width=180mm]{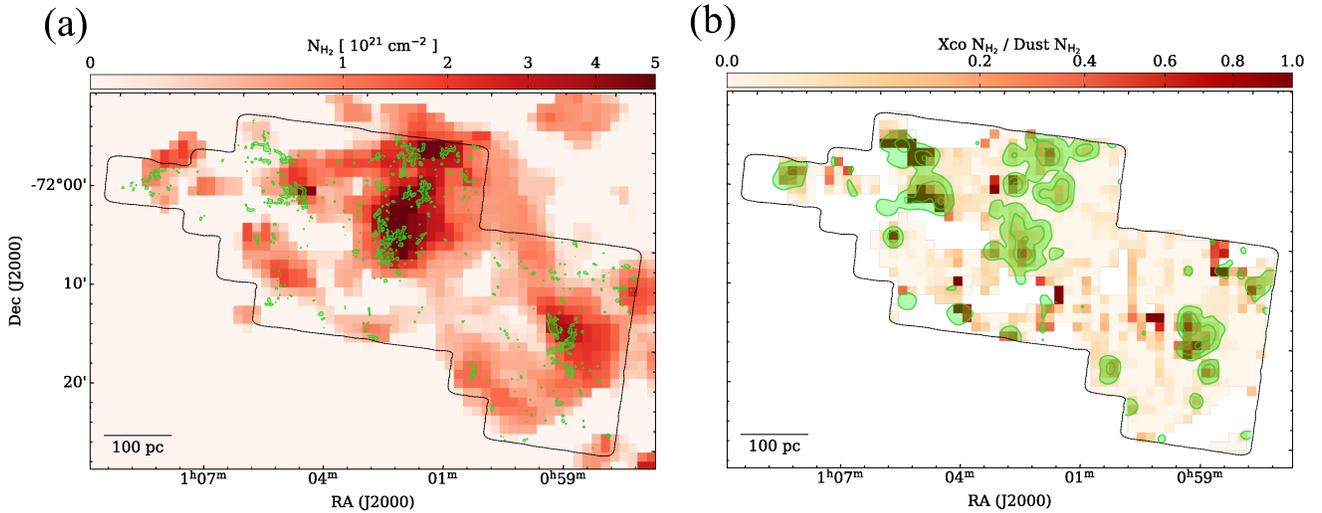}
\caption{(a) The color-scale image shows the H$_2$ column density map \citep{Jameson16}. The green contours show the moment~0 map of $^{12}$CO($J$= 2--1) with the lowest contour level of 0.5\,K\,km\,s$^{-1}$. (b) The ratio map of H$_2$ column densities derived from $X_{\rm CO}$ based analysis and that in panel (a) (see the text in Sect.~\ref{dis:COprop}). The green contours are the same as in panel (a) but for spatially smoothed data with a beam size of 1$\arcmin$ with the contour levels of [0.05, 0.5, 1.5] K\,km\,s$^{-1}$.
\label{fig:nh2}}
\end{figure}

Based on the multi-wavelength Herschel PACS/SPIRE data, \cite{Jameson16} estimated column density distributions of molecular hydrogen in the SMC (hereafter, dust $N_{\rm H_2}$) at a spatial resolution of $\sim$10\,pc by subtracting the contribution of atomic (H$\;${\sc i}) gas. Figure~\ref{fig:nh2}(a) overlays the CO contours at the native spatial resolution of $\sim$2\,pc on the dust $N_{\rm H_2}$ map. The column density at the compact CO cloud positions ranges $\sim$(1--10) $\times$10$^{20}$\,cm$^{-2}$. Their CO integrated intensity at $\sim$2\,pc resolution is typically $\sim$1\,K\,km\,s$^{-1}$, corresponding to H$_2$ column density of $\sim$6\,$\times$10$^{20}$cm$^{-2}$ (see the assumptions in Sect.~\ref{dis:Xco}). This comparison indicates that high-density compact CO clouds are locally embedded at the diffuse H$_2$ molecular clouds that are not necessarily traced by CO emission.  

We tentatively derive the total amount of CO-dark molecular clouds by comparing our CO-based mass and the dust emission.
Figure\,\ref{fig:nh2}(a) shows the dust $N_{\rm H_2}$ map derived by \cite{Jameson16}, which appears to be more widely spread than the CO contours. To compare the two different maps more quantitively, we spatially smoothed the CO-based $N_{\rm H_2}$ map to be the same spatial resolution ($\sim$10\,pc) as the dust $N_{\rm H_2}$ map and then made the ratio map between the two independent measurements (Figure\,\ref{fig:nh2}(b)).   
Although we see the ratio of $\sim$1\ at some CO strong spots, most of the observed field does not exceed the value of 0.2. 
We derived total H$_2$ mass as $\sim$6$^{+6}_{-3}\times$10$^6$\,$M_{\odot}$ within the ACA observed field using the dust-based $N_{\rm H_2}$ map (see \cite{Jameson16} regarding the description of the factor of two uncertainty). 
The total CO-based molecular mass is $\sim$4\,$\times$10$^{5}$\,$M_{\odot}$, and thus we cannot see more than $\sim$90\% molecular material in CO, at least in this observed field. This value is significantly higher than the MW observation, $\sim$30\% \citep{Grenier05}, and modeling/numerical studies to mimic MW-like conditions \citep{Wolfire10,Smith14}. Note that there is a caveat in this simple estimation of CO-dark H$_2$ gas because some studies in the MW suggest that the presence of dark gas is alternatively explained as optically thick cold H$\;${\sc i} gas \citep[e.g.,][]{Fukui15b,Hayashi19}. Higher dust $N_{\rm H_2}$ regions without any CO detection (Figure~\ref{fig:nh2}) may represent that the H$\;${\sc i} emission is saturated due to a high opacity and the analysis failed to fully subtract the atomic gas contribution from the total hydrogen column density. On the other hand, \cite{Jameson19} reported that there is not a significant component of optically thick H$\;${\sc i} gas based on their absorption measurements toward 37 out of a total of 55 detected background continuum sources. In this case, the large spatial discrepancy between the dust $N_{\rm H_2}$ and ACA CO maps may be likely due to the lower resolution H$\;${\sc i} data, $\sim$90\arcsec, than the Herschel-based dust data, $\sim$50\arcsec. Although it is beyond this paper's scope to coordinate the controversial results \cite[c.f.,][]{Murray18}, the detailed analysis using other wavelength data such as H$\;${\sc i} and gamma-ray emission are desired to understand further the large gap between the CO and dust emission in the SMC. 

\section{Summary} \label{sec:sum}
We presented the CO($J$ = 2--1) survey in the SMC North region obtained by the ACA stand-alone mode. The field coverage and the beam size are 0.26\,degree$^{2}$ ($\sim$2.9 $\times$ 10$^{5}$ \,pc$^2$) and 6$\arcsec$ ($\sim$2\,pc), respectively. The survey qualities and our early analysis can be summarized as follows:

\begin{enumerate}
  \item The detection limit in CO luminosity ($L_{\rm CO(2-1)}$) is $\sim$1.0\,K\,km\,s$^{-1}$\,pc$^{2}$, corresponding to the mass detection threshold of $\sim$10--20\,$M_{\odot}$ with assumptions of $\alpha_{\rm CO}$ = 13\,$M_{\odot}$\,(K\,km\,s$^{-1}$\,pc$^{2}$)$^{-1}$ and $R_{\rm 2-1/1-0}$ = 0.9. This sensitivity is two orders of magnitude higher than that of the previous complete CO survey in the SMC \citep{Mizuno01}.   
  \item The previously known CO clouds are resolved into spatially-isolated clustered clumps rather than single giant molecular clouds. The sensitive survey detects new faint CO emission (the compact CO clouds) at the positions down to lower H$_2$ column density ($\sim$10$^{20}$\,cm$^{-2}$) region, judging from the Herschel measurement at ten pc resolution. The observed clouds have a typical peak brightness temperature of $\lesssim$1\,K. The possible interpretation is that we cannot fully resolve them even with the $\sim$2\,pc resolution data. It is likely that the beam dilution effect reduces the observed temperature. 
  \item We investigated some of the compact CO clouds and found infrared point sources, including YSO candidates, associations. The good spatial correspondence to cloud peaks indicates that (high-mass) star formation is ongoing with an early phase before their parental clouds' dissipation. Follow-up studies using higher-resolution molecular line data and more diffuse gas tracers are necessary to unveil the nature of the star-forming compact clouds and accretion process onto the inside protostars. 
  \item At least within the observed field in the SMC, one of the notable characteristics is that more than 90\% of the molecular material is not traced by CO, which is significantly higher than that of the MW study.

\end{enumerate}

\acknowledgments

We would like to thank the referee, Dr. Katherine E. Jameson for useful comments that improved the manuscript.
This paper makes use of the following ALMA data: ADS/ JAO. ALMA\#2017.A.00054.S, and \#2015.1.01296.S. ALMA is a partnership of ESO (representing its member states), NSF (USA) and NINS (Japan), together with NRC (Canada), MOST and ASIAA (Taiwan), and KASI (Republic of Korea), in cooperation with the Republic of Chile. The Joint ALMA Observatory is operated by ESO, AUI/NRAO, and NAOJ. This work was supported by NAOJ ALMA Scientific Research grant Nos. 2016-03B and Grants-in-Aid for Scientific Research (KAKENHI) of Japan Society for the Promotion of Science (JSPS; grant Nos. JP18K13582, JP18H05440, and JP21H00049).  T.W. acknowledges support from NSF grant AST-2009849.  We thank Dr. Kei E.I. Tanaka and Dr. Thomas G. Bisbas for discussions on the CO cloud properties and line ratio from theoretical aspects.
\software{CASA (v5.4.0; \citealt{McMullin07}), Astropy \citep{Astropy18}, APLpy \citep{Robi12}}

\appendix \label{App}

We provide CO($J$ = 2--1) and CO($J$ = 1--0) data cube and 2-D continuum images in the FITS format as online materials\footnote{https://doi.org/10.5281/zenodo.4628967}. This Appendix presents the data quality/continuum maps and the intensity ratio of the CO lines.  

\section{Band~6/3 data: noise/continuum maps and flux comparison between the CO(2-1) 7\,m and TP array data} \label{Ap:RMScont}
\begin{figure}[htbp]
\includegraphics[width=180mm]{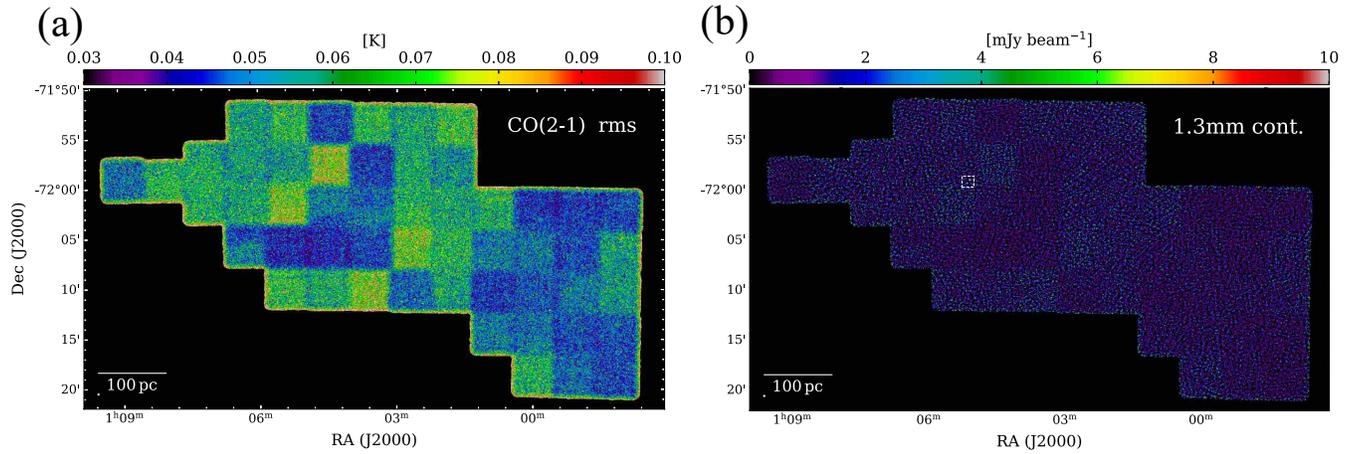}
\caption{(a) The noise level map of CO($J$ = 2--1) data in Band6. (b) The 1.3\,mm continuum map. The white rectangle shows the displayed region in Figure~\ref{fig:N78_cont} (b,c).
\label{fig:Band6noise}}
\end{figure}

\begin{figure}[htbp]
\includegraphics[width=180mm]{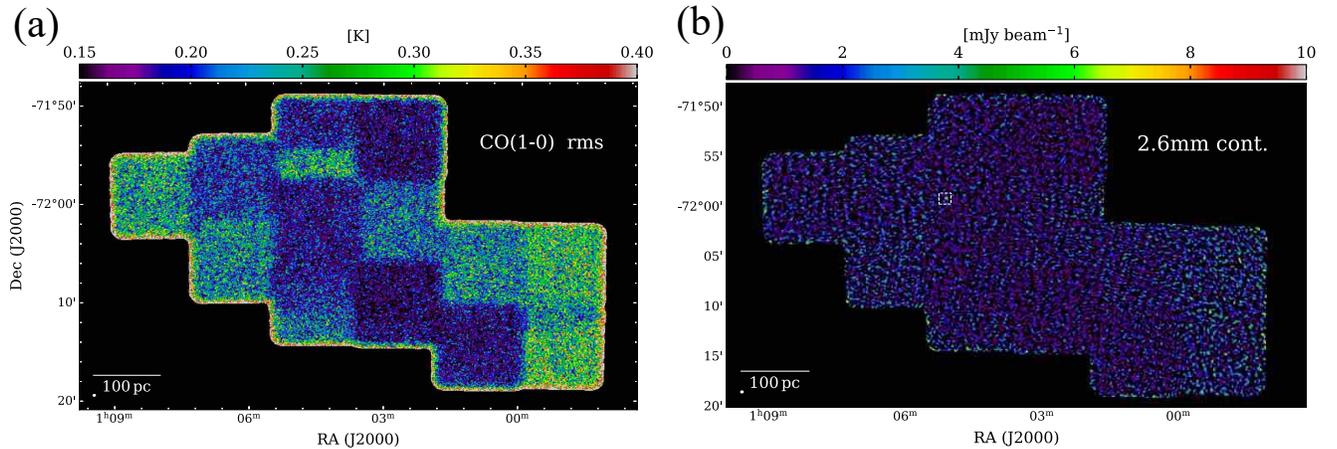}
\caption{Same as Figure~\ref{fig:Band6noise} but for Band~3.
\label{fig:Band3noise}}
\end{figure}

Figures~\ref{fig:Band6noise} and \ref{fig:Band3noise} show the r.m.s. noise level of the CO data and 1.3/2.6\,mm continuum maps. The noise levels are somewhat depending on region to region, 0.02--0.07\,K for CO($J$ = 2--1) and 0.2--0.35\,K for CO($J$ = 1--0), due to different observing conditions in each sub-map. 

\begin{figure}[htbp]
\includegraphics[width=180mm]{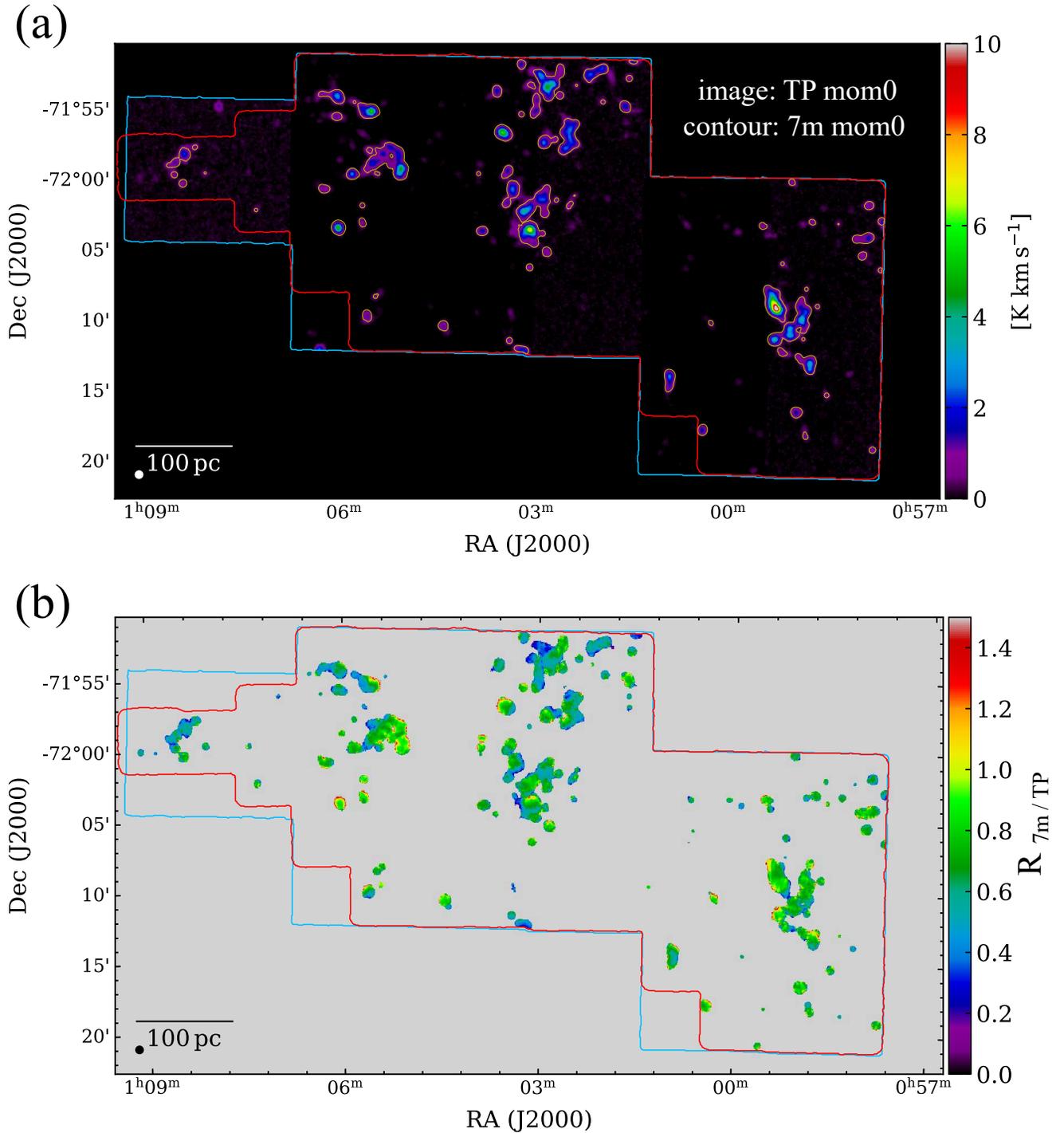}
\caption{(a) The color-scale image illustrates the TP moment~0 map in CO(2--1). Yellow contours show the smoothed moment~0 map obtained with the 7\,m with a contour level of 0.3\,K\,km\,s$^{-1}$. The white circle at the lower left corner is the TP array beam size, 30$\arcsec$. The red and blue lines show the field coverage of 7\,m and TP, respectively. (b) The color-scale image shows distributions of the ACA 7\,m/TP CO($J$ = 2--1) integrated intensity ratio. We ignored a low level emission less than 0.06\,K, which corresponds to 3$\sigma$ level of the TP array data. The integrated velocity range is 114.5--232.0\,km\,s$^{-1}$. The black circle at the lower left corner is the TP array beam size, 30$\arcsec$.
\label{fig:7mTPratio}}
\end{figure}

Figure~\ref{fig:7mTPratio} shows the TP array and spatially smoothed ACA 7\,m array data images in CO(2--1). We coordinated the angular resolutions of both data as 30$\arcsec$, and then, we made the intensity ratio map (panel (b)). The spatially smoothed 7\,m array data well reproduce the TP array distributions, and the mean flux ratio is $\sim$0.7, indicating that more than half of the total CO flux is recovered by the 7\,m array alone.

\section{CO($J$ = 1--0) map and line ratio}\label{Ap:COratio}

\begin{figure}[htbp]
\centering
\includegraphics[width=150mm]{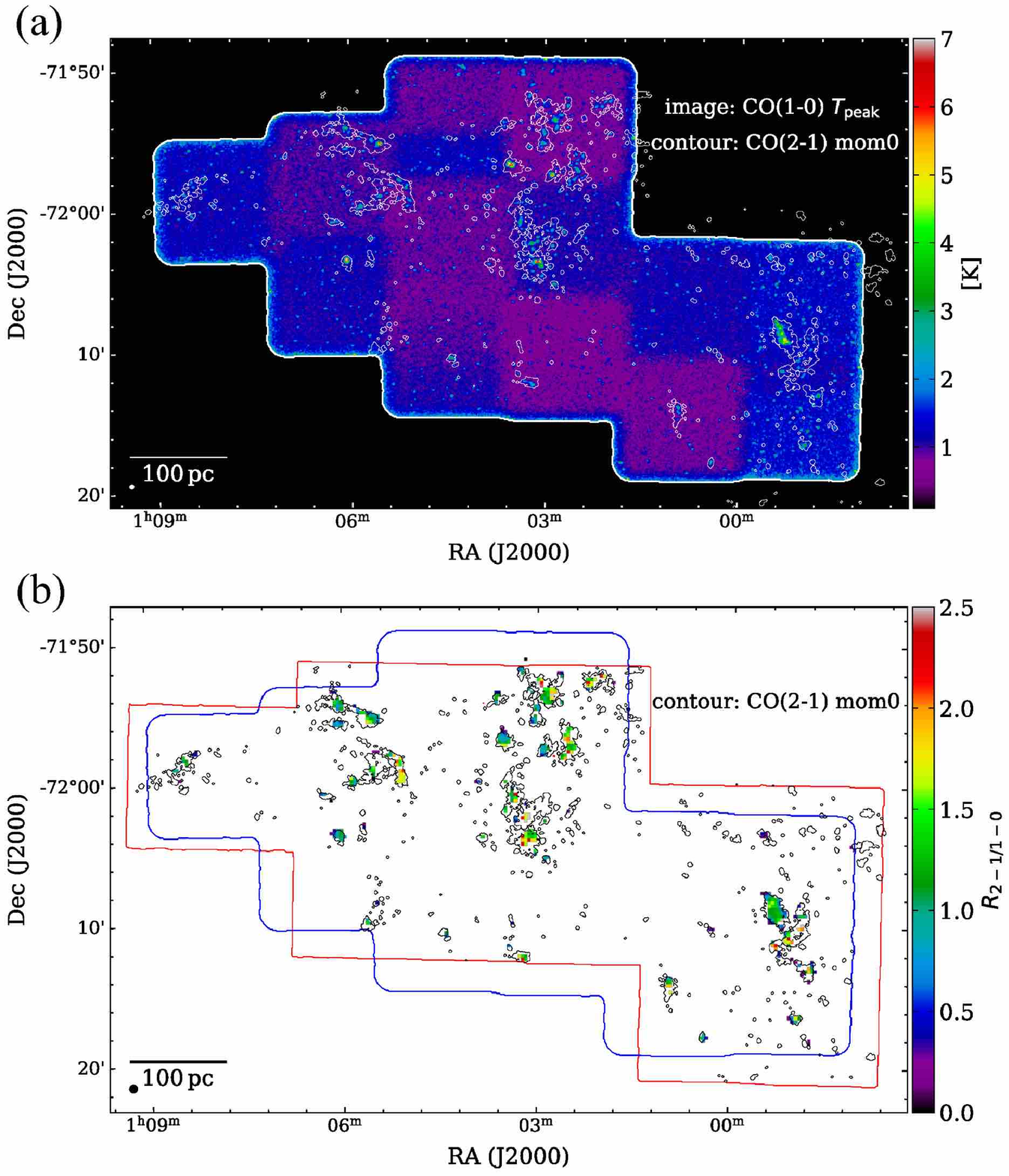}
\caption{(a) The color-scale image illustrates the CO($J$ = 1--0) $T_{\rm peak}$ map. The beam size is given by the white ellipse in the lower-left corner. White contours show the moment~0 map of CO($J$ = 2--1) with the contour level of 1\,K\,km\,s$^{-1}$. (b) Integrated intensity ratio, CO($J$ = 2--1)/CO($J$ = 1--0) ($R_{\rm 2-1/1-0}$) map (see the text). The smoothed beam size (30$\arcsec$) is shown in the lower-left corner. Black contours are the same as the white ones in panel (b). The red and blue lines show the field coverage of CO($J$ = 2--1) and CO($J$ = 1--0), respectively. \label{fig:COratio}}
\end{figure}

Figure\,\ref{fig:COratio}(a) illustrates the CO($J$ = 1--0) distribution in peak brightness temperature, $T_{\rm peak}$. Due to the poor sensitivity and the beam dilution effect, most of the compact features in CO($J$ = 2--1), whose size is close to the Band~6 beam element with the intensity of $\sim$1\,K, are not detected in CO($J$ = 1--0). Since this project did not include TP array observations, we cannot precisely estimate the 7\,m array data's missing flux. The spatially smoothed 7\,m array data with a beam size of 2\farcm6 reproduces the overall CO distribution revealed by the NANTEN survey \citep{Mizuno01}. \cite{Muller10} observed several regions in the SMC North region with the single-dish Mopra telescope whose beam size is $\sim$42$\arcsec$. The CO intensities of the 7\,m array corresponds to those with Mopra (see Table~2 in \citealt{Muller10}) within the measurement error at least in the available CO intense spots. These comparisons tell us that the missing flux of the 7\,m array CO($J$ = 1--0) data is not serious.

\begin{figure}[htbp]
\centering
\includegraphics[width=130mm]{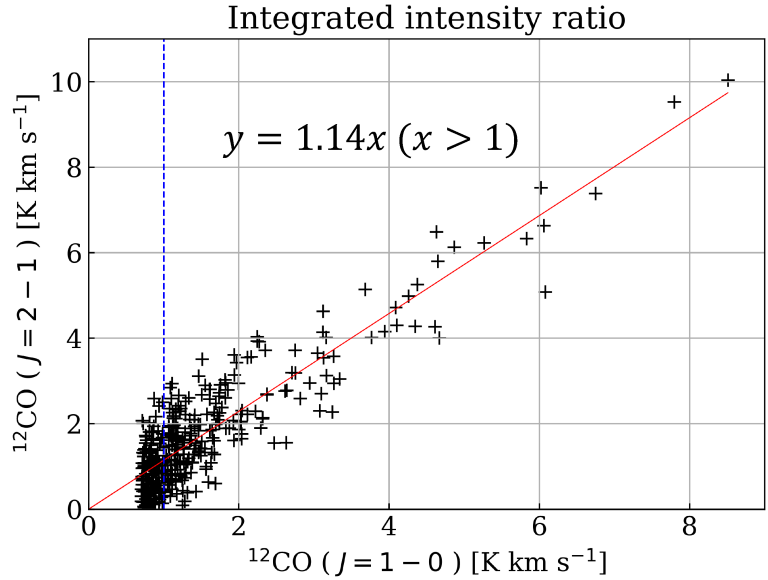}
\caption{The correlation plot between the integrated intensity of CO($J$ = 2--1) and CO($J$ = 1--0) at an angular resolution of 30$\arcsec$. The red line shows the fitting result obtained by the least-square method. The weak emission less than 1\,K\,km\,s$^{-1}$ (shown in the blue dotted line) were not used in the fitting. 
\label{fig:ratio_plot}}
\end{figure}

To discuss the overall intensity ratio of the two transitions ($J$ = 2--1 and 1--0), $R_{\rm 2-1/1-0}$, we spatially smoothed both CO data into the beam size of $\sim$30$\arcsec$. We determined the integrated velocity range based on the higher sensitivity 2--1 data, and then we applied the same velocity range to make the CO($J$ = 1--0) moment~0 map. Figure~\ref{fig:COratio}(b) and Figure~\ref{fig:ratio_plot} show the integrated intensity ratio map and the correlation plot, respectively. Note that the current analysis does not contain outer layers of the CO clouds due to the poor sensitivity of the 1--0 emission. The least-square fitting tells us that the slope is 1.1. Since the CO($J$ = 1--0) presumably underestimate the flux due to the interferometric effect (but it is not very large), we adapt $R_{\rm 2-1/1-0}$ ratio of 0.9 as a conservative way to estimate the CO luminosity-based mass (see Sect.~\ref{dis:COprop}).

\bibliography{sample63}{}
\bibliographystyle{aasjournal}

\end{document}